\documentclass[conference]{IEEEtran}
\IEEEoverridecommandlockouts
\usepackage{cite}
\usepackage{amsmath,amssymb,amsfonts}
\usepackage{algorithmic}
\usepackage{graphicx}
\usepackage{textcomp}
\def\BibTeX{{\rm B\kern-.05em{\sc i\kern-.025em b}\kern-.08em
    T\kern-.1667em\lower.7ex\hbox{E}\kern-.125emX}}
\usepackage{subcaption}
\usepackage{xspace}
\usepackage{multirow}
\usepackage{array}
\usepackage[dvipsnames,table,xcdraw]{xcolor}
\usepackage[hidelinks]{hyperref}
\usepackage{url}  
\usepackage{breakurl}

\hypersetup{
    colorlinks,
    linkcolor={[rgb]{0,0,1}},
    citecolor={[rgb]{0,0,1}},
    urlcolor={[rgb]{0,0,1}},
    breaklinks=true,
}

\usepackage{tikz}
\usetikzlibrary{positioning, intersections, arrows, shapes.geometric, calc,
  shapes.callouts, decorations.pathreplacing, decorations.markings, shapes, patterns}
\usepackage{pifont}
\usepackage{pgfplots}
\usepgfplotslibrary{statistics}
\usepackage{pgfplotstable}
\pgfplotsset{compat=newest}
\usepackage{filecontents}
\usepackage[utf8x]{inputenc}
\usepackage{lipsum}
\usepackage{soul}
\usepackage{comment}
\usepackage{footnote}
\usepackage{siunitx}
\usepackage{etoolbox}
\usepackage{soulpos}
\usepackage{wrapfig}
\usepackage{siunitx}

\usepackage{atbegshi}
\usepackage{fancyhdr}

\fancypagestyle{firstpage}{
    \fancyhf{}
    \fancyhead[C]{\vspace{15pt}\normalsize{Preprint Version. Final version to appear in IEEE International Conference on Cloud Computing, 2021}} 
    \fancyfoot[C]{\thepage}
}

\makesavenoteenv{tabular}
\makesavenoteenv{table}

\def\BibTeX{{\rm B\kern-.05em{\sc i\kern-.025em b}\kern-.08em
    T\kern-.1667em\lower.7ex\hbox{E}\kern-.125emX}}

\ulposdef{\gray}[xoffset=1pt]{\mbox{\color{gray!30}\rule[-.8ex]{\ulwidth}{3ex}}}

\newcommand\TPU{EdgeTPU\xspace}
\newcommand\TPUs{EdgeTPUs\xspace}
\newcommand\DevB{DevBoard\xspace}
\newcommand\USBA{USB-Accelerator\xspace}
\newcommand\USBAs{USB-Accelerators\xspace}
\newcommand\TX{J.TX2\xspace}
\newcommand\Nano{J.Nano\xspace}
\newcommand\ODN{ODN2\xspace}
\newcommand\RPI{RPi4\xspace}

\newcommand\alexnet{AlexNet\xspace}
\newcommand\densenet{DenseNet-161\xspace}

\newcommand\incept{Inception-V3\xspace}
\newcommand\mobone{MobileNet-V1\xspace}
\newcommand\mobtwo{MobileNet-V2\xspace}
\newcommand\mobboth{MobileNet-V1/V2\xspace}
\newcommand\resneteighteen{ResNet-18\xspace}
\newcommand\resnetfifty{ResNet-50\xspace}

\newcommand\squeezenet{SqueezeNet-V1\xspace}
\newcommand\vgg{VGG-16\xspace}

\def\footnoterule{\relax%
  \kern0pt
  \hbox to \columnwidth{\hfill\vrule width 1.0\columnwidth height 0.4pt\hfill}
  \kern4.6pt}

\begin{document}

\title{\huge AI Multi-Tenancy on Edge: Concurrent Deep Learning Model Executions and Dynamic Model Placements on Edge Devices}

\author{\IEEEauthorblockN{
	Piyush Subedi,
    Jianwei Hao,
    In Kee Kim,
    Lakshmish Ramaswamy}
    \IEEEauthorblockA{Department of Computer Science, University of Georgia, \{piyush.subedi, jhao, inkee.kim, laksmr\}@uga.edu}
}

\maketitle
\thispagestyle{firstpage}
\pagestyle{plain}

\begin{abstract}
Many real-world applications are widely adopting the edge computing paradigm due to its low latency and better privacy protection. With notable success in AI and deep learning (DL), edge devices and AI accelerators play a crucial role in deploying DL inference services at the edge of the Internet. While prior works quantified various edge devices' efficiency, most studies focused on the performance of edge devices with single DL tasks. 
Therefore, there is an urgent need to investigate AI multi-tenancy on edge devices, required by many advanced DL applications for edge computing.

This work investigates two techniques -- {\em concurrent model executions} and {\em dynamic model placements} -- for AI multi-tenancy on edge devices. 
With image classification as an example scenario, we empirically evaluate AI multi-tenancy on various edge devices, AI accelerators, and DL frameworks to identify its benefits and limitations. 
Our results show that multi-tenancy significantly improves DL inference throughput by up to 3.3$\times$ -- 3.8$\times$ on Jetson TX2. 
These AI multi-tenancy techniques also open up new opportunities for flexible deployment of multiple DL services on edge devices and AI accelerators.

\end{abstract}


\section{Introduction}\label{sec:intro}
There have been massive strides in Artificial Intelligence (AI) and Deep Learning (DL) technologies in recent years. Newer DL algorithms coupled with highly cost-effective and scalable mechanisms to gather, store, and process large amounts of data have led to what some researchers believe to be the golden age of AI/DL~\cite{bughin-WP-2017}. It is widely expected that in the near future, AI will drive applications in many massively distributed domains, such as autonomous vehicles, disaster response, precision agriculture, and drone-based surveillance~\cite{EdgeIntell2019, edge-autonomous-2019}. These domains are often distinguished by two fundamental characteristics, namely, stringent response time requirements (i.e., real-time or near real-time), data sources that are distributed at the edge of the Internet and highly resource-constrained operational environments~\cite{Serving@theEdge2017}.

The predominant paradigm for building AI systems is to centralize all AI tasks at the cloud~\cite{GCE-AI:online2019, AWS-AI:online2019, Azure-AI:online2019, IBM-AI:online2019, stoica-CORR-2017}. In other words, in this {\em cloud-based AI paradigm}, pre-trained (often large-scale) DL models are deployed exclusively at the cloud~\cite{kang-asplos-17, Hadidi-RAL-2018, Jeong-SoCC-2018, Talagala-HotEdge-2018, Zhang-HotEdge-2018, Chinchali-HotNet-2018}. 
While cloud-based AI offers distinct advantages, particularly to domains, e.g., social networks, e-commerce, and finance, where the data is naturally available on the cloud, this paradigm is not well suited for the aforementioned domains. 
This is because transferring large amounts of data from the network edge to the cloud over low-bandwidth connections is prohibitively expensive, often resulting in AI service disruptions caused by network disconnections, which are not well tolerated by these applications. 

Towards addressing the above limitations, researchers have recently been exploring the \emph{AI at the edge} paradigm~\cite{DLwithEdge-PIEEE-2019, EdgeIntell2019, AI-on-edge-cacm, EdgeAI-TWC-2020, SNTA21:Characterization}, where
DL applications are hosted at the {\em edge of the Internet} (e.g., closer to the data sources). Advent and proliferation of miniaturized yet powerful computing boards, e.g., Raspberry Pi~\cite{RPI4}, Nvidia Jetson Nano~\cite{JetsonNano}, and Google's \TPU~\cite{EdgeTPU}, have served as key enablers for moving DL tasks to the edge of the Internet. Several studies have been conducted to quantify the efficiency of various edge devices for DL inference tasks~\cite{pCAMP-HotEdge-2018, EmBench-CoRR19, ResCharct-AUChallengeIoT19, PerfUSBAccelerator-2020, CharacterizingDNNonEdge-IIWSC2019, AIEdge-CoRR-2020}. Most existing studies have focused on characterizing the performance (e.g., latency and throughput) of edge devices and AI accelerators with single DL tasks. 
However, many advanced applications often require AI multi-tenancy where multiple DL tasks are co-running on edge devices. For instance, drone-based surveillance requires simultaneous executions of inference tasks on video and audio streams~\cite{DroneExample2019}. Unfortunately, very few researchers have tried to quantify and optimize the AI multi-tenancy on edge devices to the best of our knowledge.

Leveraging AI multi-tenancy on edge devices has the potential to provide distinct benefits in offering DL services. 
AI multi-tenancy can be achieved via leveraging {\em concurrent model executions} (CMEs) and {\em dynamic model placements} (DMPs).
CME allows the deployment of multiple DL models on either GPU or \TPU resources and runs them in parallel.
Thus, CME can potentially improve the overall DL inference throughput and enable the execution of different DL applications/models simultaneously. 
DMP enables AI multi-tenancy by deploying and executing DL models on different resources on edge devices at the same time. e.g., DL models on both GPU and \TPU. 
DMP is particularly useful when AI accelerators (e.g., EdgeTPU) enhance edge devices, and it can significantly increase the resource utilization and the DL inference throughput by utilizing multiple resources on the devices and the accelerators.

While there are expected advantages of AI multi-tenancy on edge devices, it is also important to identify the limitations of AI multi-tenancy to maximize the benefits of AI at the edge. 
Specifically, in this work, we seek answers to the following research questions. 
{\em What are the performance benefits of enabling AI multi-tenancy in the device level?}
{\em What are the limitations of the edge devices and accelerators to support AI multi-tenancy, such as the limit of model concurrency, resource contention, and resource bottleneck?}

To answer the above research questions, this study performs comprehensive evaluations of CME and DMP for AI multi-tenancy and discovers
the opportunities and limitations of such approaches. 
Both CME and DMP are thoroughly evaluated with widely used edge devices and \TPU accelerators. 
We use {\em image classification} as an example application scenario of AI at the edge and assess nine pre-trained DL models\footnote{These DL models are pre-trained models of CNN (Convolutional Neural Network) models for image classifications.} with four DL frameworks.

We first characterize the behavior and performance (e.g., inference throughput) of both edge devices and \TPU accelerators and identify critical resource factors affecting the DL inference throughput on the edge devices and accelerators. 
Then we apply two AI multi-tenancy approaches to DL inference tasks on the devices and accelerators, and then we discover the empirical upper bound of DL inference throughput as well as the impact from resource contention.
Our evaluation results show that modern edge devices and \TPUs can achieve 1.9$\times$ -- 3.3$\times$ higher inference throughput with CME. Moreover, the DMP approach can increase throughput by up to 3.8$\times$. 
These two approaches for AI multi-tenancy open up new opportunities for maximizing the resource utilization of devices and flexible deployment of multiple DL applications on edge devices. 

The research contributions of this work are as follows:

1. We thoroughly characterize and quantify the performances (DL inference throughput) and behaviors of various edge devices and AI accelerators when enabling AI multi-tenancy. Such characterizations are performed by employing a set of DL frameworks and DL models widely used for image classifications.

2. We discover the empirical upper bound of the performance and the model concurrency on edge devices and \TPUs when AI multi-tenancy is enabled by CME. 

3. We identify the performance benefits and limitations when adopting DMP to utilize heterogeneous resources on edge resources and \TPUs. 
This work is the first study to characterize and evaluate DMP for AI multi-tenancy to the best of our knowledge.

We structure the rest of the paper as follows. 
Section~\ref{sec:background} describes edge devices, \TPUs, DL models, and DL frameworks used in this work. 
Section~\ref{sec:single-tenancy} characterizes the performance and the behavior of the devices with single-tenancy cases.
Section~\ref{sec:multi-tenancy} conducts evaluations of two AI-multi-tenancy techniques on edge devices and AI accelerators and describes their benefits and limitations.
Section~\ref{sec:discussion} summarizes and discusses our findings from this work.
Section~\ref{sec:related} describes related work, and  Section~\ref{sec:conclusion} concludes this paper.

\section{Edge Devices, \TPU Accelerators, Deep Learning Models, and Deep Learning Frameworks}\label{sec:background}
This section describes edge devices and AI accelerators, DL models, and DL frameworks used in this study.

\subsection{Edge Devices and \TPU Accelerators}\label{subsec:devices}

In this work, we employed the following four widely-used edge devices and two \TPU AI accelerators.

\vspace{0.75mm}
\noindent
{\bf Jetson TX2 (\TX)~\cite{JetsonTX2}} is a high-performance edge device with six CPU cores (a dual-core Denver 2 CPU and a quad-core ARM Cortex-A57 at 2 GHz) and a 256-core Nvidia Pascal GPU for DL processing. \TX has a 8 GB LPDDR4 RAM, which is shared by CPUs and GPUs.
Among five different power modes in \TX~\cite{TX2-PowerMode}, we use {\tt mode-0} ({MaxN}) to enable all six CPU cores and provide the highest frequency of both CPUs (2.0 GHz) and GPUs (1.3 GHz). 

\vspace{0.75mm}
\noindent
{\bf Jetson Nano (\Nano)~\cite{JetsonNano}} is a small yet powerful single board computer specialized in DL processing. It has a quad-core ARM Cortex-A57 (1.5 GHz), a 128-core Nvidia Maxwell GPU, and 4 GB LPDDR4 RAM (shared by both CPUs and GPUs). For \Nano, we use a power mode of {\tt mode-0}, which is default mode for maximizing the device performance.

\vspace{0.75mm}
\noindent
{\bf Odroid-N2 (\ODN)~\cite{ODN2}} is a computing board with 4GB LPDDR4 RAM and six CPU cores (a quad-core Cortex-A73 at 1.8 GHz and dual-core Cortex-A53 at 1.9 GHz). 
While \ODN has a GPU (Mali-G52 GPU), we cannot use this GPU for DL inference tasks due to a software compatibility issue. 

\vspace{0.75mm}
\noindent
{\bf Raspberry Pi 4 (\RPI)~\cite{RPI4}} is a small, low-cost, representative computing board for edge/IoT devices. \RPI is based on Broadcom BCM2711 SoC and has a quad-core ARM Cortex-A72 (1.5 GHz) and 4 GB LPDDR4 RAM. \RPI neither has a GPU nor specialized HW accelerators for DL processing.

\vspace{0.75mm}
\noindent
{\bf Coral Dev Board (\DevB)~\cite{DevBoard}} is a single-board computer equipped with a quad-core Cortex-A53 CPU (1.5GHz) and 1GB LPDDR4 RAM\footnote{The newer version of \DevB can have 2G or 4G of LPDDR4 RAM, but we use \DevB with 1GB RAM.}, as well as onboard TPU (Tensor Processor Unit) co-processor, performing 4 trillion operations per second (TOPS) at 2W of power consumption.

\vspace{0.75mm}
\noindent
{\bf Coral USB Accelerator (\USBA)~\cite{USBAcc}} 
is a USB-type TPU accelerator for machine learning (ML) and DL. The performance of its onboard \TPU accelerator is equivalent (4 TOPS at 2W) to that in \DevB. \USBA can be connected with diverse host edge devices (e.g., \RPI and \Nano) and enhance DL processing.
Since it only has an \TPU co-processor, \USBA relies on the host device's memory system to store and load the DL models and their parameters.

\subsection{DL Models, Frameworks, and Application}\label{subsec:CNN_models}

\begin{table*}[t]
    \caption{The overview of 9 DL models}
    \vspace{-2mm}
    \centering
    \def\arraystretch{1.5}%
    \begin{tabular}{
    |l|c|c|c|c|c|c|c|c|c|c|} 
        \hline
    \multicolumn{1}{|l|}{\multirow{2}{*}{}} & \multicolumn{1}{c|}{\multirow{2}{*}{\bf Year}} & \multicolumn{1}{c|}{\multirow{2}{*}{\begin{tabular}[c]{@{}c@{}}\bf Input\\ \bf Size\end{tabular}}} & \multicolumn{1}{c|}{\multirow{2}{*}{\begin{tabular}[c]{@{}c@{}}\bf Num.\\ \bf Layers\end{tabular}}} & \multicolumn{1}{c|}{\multirow{2}{*}{\begin{tabular}[c]{@{}c@{}}\bf Billion\\ \bf FLOPS\end{tabular}}} & \multicolumn{1}{c|}{\multirow{2}{*}{\begin{tabular}[c]{@{}c@{}}\# \bf Params\\ \bf (Million)\end{tabular}}} & \multicolumn{1}{c|}{\multirow{2}{*}{\begin{tabular}[c]{@{}c@{}}\bf Approx.\\ \bf File Size (MB)\end{tabular}}} & \multicolumn{4}{c|}{\bf DL Framework (FW) Support} \\ \cline{8-11} 
    \multicolumn{1}{|c|}{} & \multicolumn{1}{l|}{} & \multicolumn{1}{l|}{} & \multicolumn{1}{c|}{} & \multicolumn{1}{l|}{} & \multicolumn{1}{c|}{} & \multicolumn{1}{l|}{} & \multicolumn{1}{c|}{\bf PyTorch} & \multicolumn{1}{l|}{\bf MXNet} & \multicolumn{1}{c|}{\bf TF} & \multicolumn{1}{l|}{\bf TFLite} \\ 
        \hline        
       \alexnet~\cite{model-alexnet} & 2012 & 224$\times$224 & 8 & 0.7 & 61 & 244 & \ding{51} & \ding{51} & \ding{51} & \ding{55} \\
       \hline
       \densenet~\cite{model-densenet} & 2016 & 224$\times$224 & 161 & 7.9 & 28.7 & 115 & \ding{51} & \ding{51} & \ding{51} & \ding{55} \\
       \hline
       \resneteighteen~\cite{model-resnet} & 2015 & 224$\times$224 & 18 & 1.8 & 11.7 & 46  & \ding{51} & \ding{51} & \ding{51} & \ding{55}\\
       \hline
       \resnetfifty~\cite{model-resnet} & 2015 & 224$\times$224 & 50 & 4.1 & 25.6 & 102 & \ding{51} & \ding{51} & \ding{51} & \ding{55}\\
       \hline
       \squeezenet~\cite{model-squeezenet} & 2016 & 224$\times$224 & 15 & 0.4 & 1.2 & 5 & \ding{51} & \ding{51} & \ding{51} & \ding{55}\\
       \hline
       \vgg~\cite{model-vgg} & 2014 & 224$\times$224 & 16 & 15.4 & 138.36 & 553 & \ding{51} & \ding{51} & \ding{51} & \ding{55}\\
       \hline
       \incept~\cite{model-inceptionv3} & 2015 & 299$\times$299 & 48 & 2.9 & 27.2 & 101, 25\textsuperscript{*} & \ding{51} & \ding{51} & \ding{51} & \ding{51} \\
       \hline
       \mobone~\cite{model-mobilev1} & 2017 & 224$\times$224 & 28 & 1.1 & 4.3 & 17, 4.5\textsuperscript{*}  & \ding{51} & \ding{51} & \ding{51} & \ding{51}\\
       \hline
       \mobtwo~\cite{model-mobilev2} & 2018 & 224$\times$224 & 20 & 0.3 & 3.5 & 14, 4\textsuperscript{*} & \ding{51} & \ding{51} & \ding{51} & \ding{51} \\
       \hline       
       \multicolumn{10}{l}{%
       \begin{minipage}{16.5cm}%
       \footnotesize \ding{51} denotes that the model runs on the DL FW, \ding{55} denotes that the model does not support the DL FW, \textsuperscript{*} means information for TFLite. 
  \end{minipage}%
}\\
    \end{tabular}

    \label{tab:CNN_models}
    \vspace{-1em}
\end{table*}

\noindent
{\bf DL Models.} This study used a set of DL models to evaluate AI multi-tenancy on edge devices and AI accelerators. 
The accuracy and the size of DL models keep increasing along with the rising complexity of model dimensions and the adding number of neural network layers. 
However, such large-size models do not fit into resource-constrained, low-capacity edge devices.
Therefore, among many available DL models, we selected nine {\em pre-trained} DL models because these models have the suitable model sizes to be deployed on the resource-constrained edge devices to perform DL inference tasks (e.g., image classifications).
Moreover, all these models have unique characteristics and behaviors, such as different network architectures, number of layers, number of parameters, and model sizes.
Such differences and the overview of the nine selected models are described in Table~\ref{tab:CNN_models}.

\vspace{0.75mm}
\noindent
{\bf DL Frameworks.}
We also used four widely-used open-source DL frameworks; PyTorch~\cite{PyTorch-NIPS2019}, MXNet~\cite{MXNet-ArXiv2015}, TensorFlow (TF)~\cite{TF-OSDI16}, and TensorFlow Lite (TFLite)~\cite{TFLite}. 
PyTorch, MXNet, and TF were used for performing CPU- and GPU-based DL inference tasks on edge devices (e.g., \TX, \Nano, \ODN, and \RPI). 
TFLite was used to run DL models on \TPU (e.g., \DevB and \USBA).

Table~\ref{tab:CNN_models} also shows DL frameworks' support for DL models. All DL models are available for PyTorch, TF, and MXNet. However, \incept, \mobone, and \mobtwo are the only DL models whose pre-trained \emph{quantized} versions are available for TFLite.

\vspace{0.75mm}
\noindent
{\bf DL Applications and Dataset.}
For the DL inference task, we used {\em image classification}, which is a common use case of computer vision and can be used as a key component in various AI applications (e.g., drone-based surveillance, hazard zone detection in autonomous driving) in edge computing~\cite{DLwithEdge-PIEEE-2019, EdgeIntell2019}. 
In an image classification task on edge, a pre-trained DL model determines text labels (e.g., dog or cat) from input image streams based on the contents. The DL models often generate multiple text labels for input images with the probabilities for images associated with a specific image class.

We used the validation dataset from ImageNet ILSVRC-2012~\cite{ILSVRC15} for input images to DL inference tasks. The validation dataset contains 50K labeled images for 1K different object categories.

\section{Evaluation of AI Single-Tenancy on Edge Devices and Accelerators}\label{sec:single-tenancy}
\graphicspath{ {./images/} }
\tikzstyle{intg}=[draw=none,minimum size=2em,text centered,text width=7.cm]

We first evaluated and characterized the performance of the edge devices and the accelerators with single-tenancy cases. The results measured in this section will be used as baselines for comparison with AI multi-tenancy cases.

\subsection{Measurement Methodology}\label{subsec:measure_method} 

To quantify the performance of single tenancy cases on the edge devices and the AI accelerators, we focused on the inference throughput of DL models as the main performance metric.
The DL inference throughput results with AI single-tenancy were measured with a set of different configurations, which were the combinations of devices, DL models, batch sizes, DL frameworks. 
Please note that we tested various batch sizes ranging from 1 to 256, but we only report the batch size resulting in the highest inference throughput.
Moreover, the maximum executable batch size could vary across different edge devices and DL models due to the limitation of devices' HW capacity (e.g., memory size) and the size of DL models.
The DL inference throughput is calculated by equation-(\ref{eq:throughput_single}). 
For the single-tenancy case, the number of inferences in equation-(\ref{eq:throughput_single}) is calculated by ``batch size'' $\times$ ``the number of batches.''
\begin{equation}
\label{eq:throughput_single}
DL~Throughput = \dfrac{Number~of~Inferences}{Total~Execution~Time}
\end{equation}

\begin{wrapfigure}{r}{0.52\columnwidth} 
     \centering
     \includegraphics[width=0.53\columnwidth]{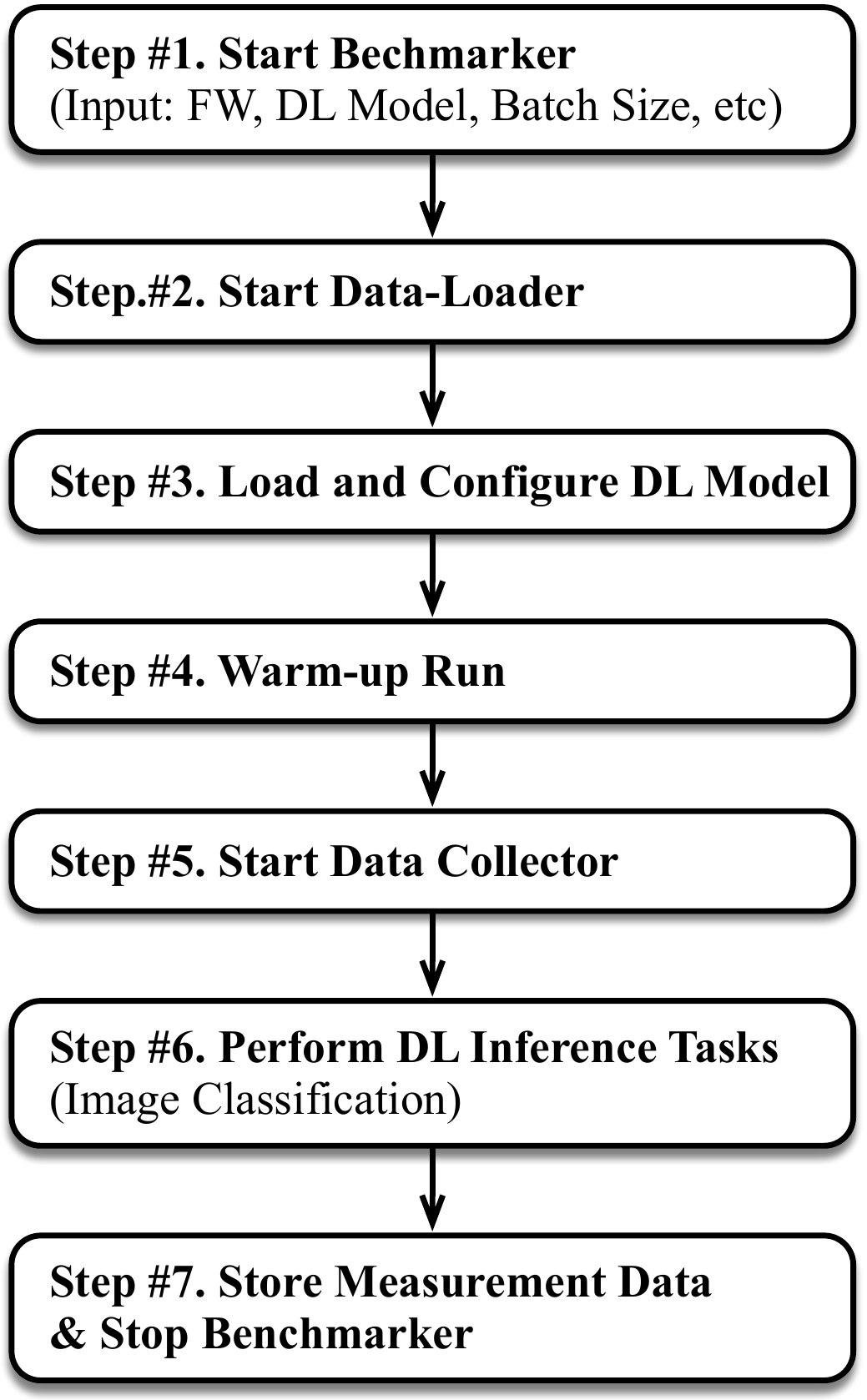}
    \caption{Measurement steps}
    \label{fig:workflow}
    \vspace{-1em}
\end{wrapfigure}

\vspace{0.75mm}
\noindent
{\bf Measurement Procedure.}

We developed a benchmarker to measure the DL inference throughput and collect necessary system statistics.
We deployed the benchmarker along with an image classification application on the devices and \TPUs accelerators.
The measurement procedure of the benchmarker is shown in Fig.~\ref{fig:workflow}. 

The benchmarker begins by taking specific parameters for the measurement (step \#1), including the DL model, DL framework, batch size, and others. 
Then, the benchmarker starts a DL framework-specific data-loader (step \#2) that prepares input images (as per the batch size) from the dataset (ImageNet ILSVRC-2012) and sends them to the DL model.
In step \#3, the benchmarker loads the DL model into the main memory and configures it based on the parameters (e.g., use of CPU, GPU, or \TPU).
The next step (step \#4) is the warm-up run phase, which ensures all the necessary components are loaded, and the DL framework configures suitable optimization strategies before performing the actual measurement.
After the warm-up run, the benchmarker starts a data collector (step \#5) that contains tools for measuring system statistics (e.g., {\tt sysstat}) and power consumption (e.g., {\tt INA-219}).
Then, in step \#6, the benchmarker performs DL inference tasks (image classification) for input images received from the data-loader. 
The inference tasks are performed at least 30 times to increase the statistical confidence of the measured data. 
While the inference tasks are performed, the data collector continuously measures resource usage and power consumption.
After completing all the inference tasks, the benchmarker saves the measured data, and it will be terminated (step \#7).

\vspace{0.75mm}
\noindent
\textbf{System Statistics and Power Measurement.}
In the above measurement step, diverse system statistics were collected while the inference tasks were being performed. {\tt sysstat}~\cite{Sysstat} was used to collect the usage of CPU, memory, and Disk/USB IO. 

For measuring the power consumption of edge devices, we used {\tt INA-219}~\cite{ina-219}, a voltage, current and power measurement chip. 
With a default resistance of \SI{0.1}{\ohm}, the chip allows measuring the power consumption with a current sensing range of $\pm$\SI{3.2}{\ampere} and a voltage range of \SI{0}{\volt} to \SI{26}{\volt}.
We used {\tt pi-ina219}~\cite{pi-ina-219}, a python library to communicate with the {\tt INA-219} chip. 
We also used {\tt jetson-stats}~\cite{jetson-stats}, a python library that provides power consumption statistics leveraging Nvidia's {\tt tegrastats} utility~\cite{tegra-stats} to measure the power consumption of \TX and \Nano. 
For \TPUs, we used a USB power meter.

\subsection{Measurement Results with Single-Tenancy}
\label{subsec:indv_throughput}

\begin{figure}[t]
    \centering
    \includegraphics[width=1\columnwidth]{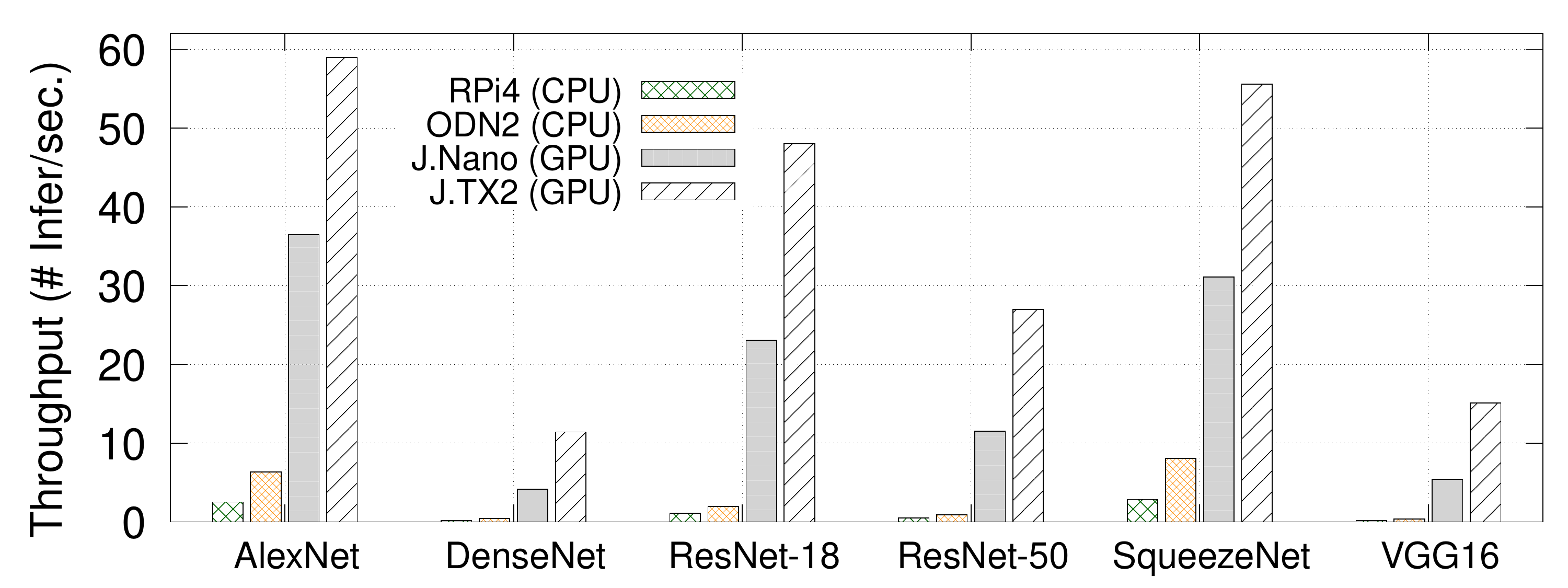}
	\caption{Inference throughput of 6 DL models on edge devices using CPU and GPU resources. Note that throughput results of CPU- and GPU-based inference results are the average of the maximum throughput results from PyTorch, TF, and MXNet.}
	\label{fig:throughput_CG}
\end{figure}

\noindent
{\bf DL Inference Throughput with Single-Tenancy.}
Fig.~\ref{fig:throughput_CG} reports the maximum DL inference throughput with single-tenancy when the DL models were executed on either CPU or GPU resources in edge devices. 
The results show that the inference throughput results varied significantly across different DL models as they had different model sizes, network architectures, and a set of parameters. 
The results also confirm that the GPU-based DL inference results showed significantly improved throughput over the CPU-based inference as GPU is more specialized in processing AI and ML workloads. 
The edge devices with GPUs (e.g., \Nano and \TX) processed 4$\times$ -- 96$\times$ more inference requests compared to the devices without GPUs (\RPI and \ODN). 
On average, \Nano showed 23$\times$ and 13$\times$ higher throughputs over \RPI and \ODN. 
\TX had 50$\times$ and 28$\times$ higher throughput results than \RPI and \ODN. 
Moreover, \TX showed 2.28$\times$ higher inference throughput than \Nano because \TX's GPU is equipped with a larger capacity GPU module (128 GPU cores in \Nano vs. 256 GPU cores in \TX).

\begin{figure}[t]
    \centering
	\includegraphics[width=1\columnwidth]{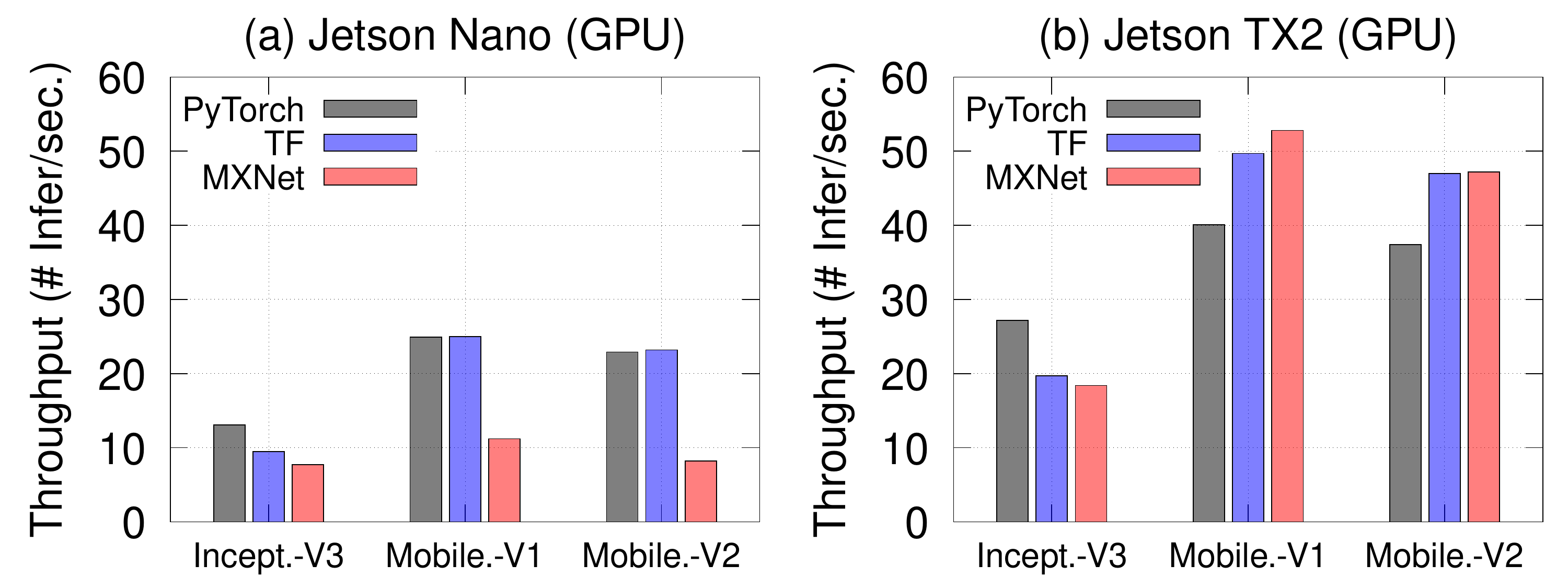}
	\caption{GPU-based inference throughput variations across three different DL frameworks. While the results report GPU-based inference throughput with three models, the other 6 DL models showed similar patterns.}
	\label{fig:gpu_fw_throughput}
\end{figure}

Moreover, we observed that the inference throughput results with GPU (e.g., \Nano and \TX) could vary significantly across three DL frameworks, as shown in Fig.~\ref{fig:gpu_fw_throughput}. In particular, MXNet on \Nano showed exceptionally (55\%) lower performance than the other two frameworks. (But GPU with MXNet on \TX did not show low inference throughput.) The lower performance with MXNet on \Nano was due to MXNet's optimization mechanism to find the best convolution algorithm for inference tasks with DL models. Unfortunately, this is a memory-intensive operation, and \Nano's 4GB memory is not large enough to complete this optimization step so that MXNet on \Nano showed poor inference throughput. 
We found the same issue in our evaluation of AI multi-tenancy when using MXNet with the CME technique.
We will provide a detailed analysis of this problem in Section~\ref{subsec:cme}.

\begin{figure}[t]
\centering
	\includegraphics[width=1\columnwidth]{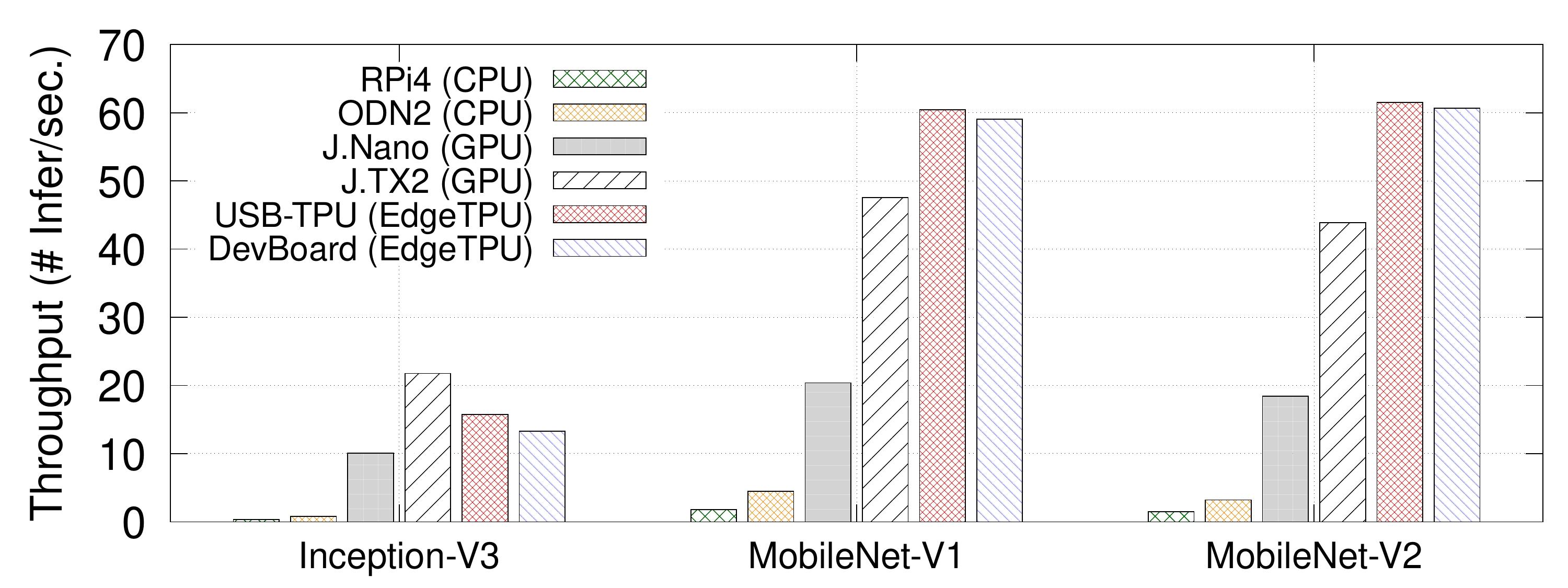}
	\caption{Inference throughput of \incept, \mobone, and \mobtwo using CPU, GPU, and \TPU. The throughput results of CPU- and GPU-based inferences are averages of maximum throughput results from PyTorch, TF, and MXNet. \USBA's throughput is the average of throughput results obtained from \USBA connected with four edge devices.}
	\label{fig:throughput_CGT}
\end{figure}

Fig.~\ref{fig:throughput_CGT} shows the comparison of maximum throughput of three DL models when they were executed on CPU, GPU, and \TPU resources.
To compute the throughput (red bar in the figure) of \USBA, we used four combinations\footnote{The four combinations are \RPI with \USBA, \ODN with \USBA, \Nano with \USBA, and \TX with \USBA.} of edge devices and \USBA, and we report the average of the maximum throughput of all four combinations. As expected, both GPU and \TPU-based inferences showed 10$\times$ -- 63$\times$ higher throughput than CPU-based inferences.
Between the GPU and \TPU resources, while \TX's 256-core Pascal GPU showed the maximum throughput (even higher than \TPU's throughput) with \incept , both \DevB and \USBA showed 25\% -- 41\% higher throughput than \TX for performing inferences with \mobboth. 

\begin{table}[t]
\caption{\TPU throughput variation across different host edge devices.}
\centering
\begin{tabular}{|l|l|c|c|}
\hline
\textbf{Model} & \textbf{Host Device + \TPU} & \textbf{Avg. Through.} & \textbf{Std. Dev.} \\ \hline
\multirow{5}{*}{\begin{tabular}[c]{@{}c@{}}Inception\\-V3\end{tabular}} & \RPI + \USBA & 12.35 & 0.35 \\ \cline{2-4} 
 & \ODN + \USBA & 15.59 & 0.47 \\ \cline{2-4} 
 & \Nano + \USBA & 16.42 & 0.34 \\ \cline{2-4} 
 & \TX + \USBA & 18.54 & 0.48 \\ \cline{2-4}
 & \DevB Only & 13.26 & 0.19 \\ \hline 
 \multirow{5}{*}{\begin{tabular}[c]{@{}c@{}}MobileNet\\-V1\end{tabular}} & \RPI + \USBA & 54.65 & 4.03 \\ \cline{2-4} 
 & \ODN + \USBA & 58.84 & 6.73 \\ \cline{2-4} 
 & \Nano + \USBA & 63.60 & 5.58 \\ \cline{2-4} 
 & \TX + \USBA & 64.65 & 5.45 \\ \cline{2-4}
 & \DevB Only & 59.02 &  2.48 \\ \hline
\multirow{5}{*}{\begin{tabular}[c]{@{}c@{}}MobileNet\\-V2\end{tabular}} & \RPI + \USBA & 55.79 & 4.15 \\ \cline{2-4} 
 & \ODN + \USBA & 59.70 & 5.78 \\ \cline{2-4} 
 & \Nano + \USBA & 66.61 & 4.23 \\ \cline{2-4} 
 & \TX + \USBA & 64.01 & 6.57  \\ \cline{2-4}
 & \DevB Only & 60.67 & 5.23 \\ \hline
\end{tabular}
\label{tab:TPU_perf_var}
\end{table}

Table~\ref{tab:TPU_perf_var} reports \TPU throughput fluctuations across different host edge devices. In particular, when performing DL inference tasks using \incept, \USBA showed up to a 33\% difference in the inference throughput on the different host devices. 
Several factors can result in such throughput fluctuations. 
Memory bandwidth on the (host) edge devices can be a factor for such fluctuations. For example, the latency when swapping in/out of a DL model and its parameters between the host devices and \USBA rely on the memory bandwidth. 
Furthermore, both storage IO and USB IO can also be factors for changing the DL inference throughput. 
Regarding these factors, we will further analyze them in the following paragraph.

\vspace{0.75mm}
\noindent
\textbf{Factors  for  Impacting  DL  Inference  Throughput Changes.}
To identify factors that change the DL inference throughput on edge devices and \TPUs, we performed correlation analysis by calculating the Pearson correlation coefficient (in equation-(\ref{eq:pcc}))~\cite{PCC:TSAL08} of measured throughput results and resource usage statistics. 
This coefficient represents the linear relationship between two variables, ranging from $-1$ to $1$. 
Please note that the coefficient of $1$ indicates an ideal positive correlation, negative values mean reverse correlation, and $0$ means there is no correlation between two variables.

\begin{equation}
\label{eq:pcc}
    \rho = \frac{cov(x,y)}{\sigma_x\sigma_y} = \frac{\sum_{i}^{n}(x_i-\overline{x})(y_i-\overline{y})}{\sqrt{\sum_{i}^{n}(x_i-\overline{x})^2(y_i-\overline{y})^2}}
\end{equation}

\begin{figure}[t]
    \centering
	\includegraphics[width=1\columnwidth]{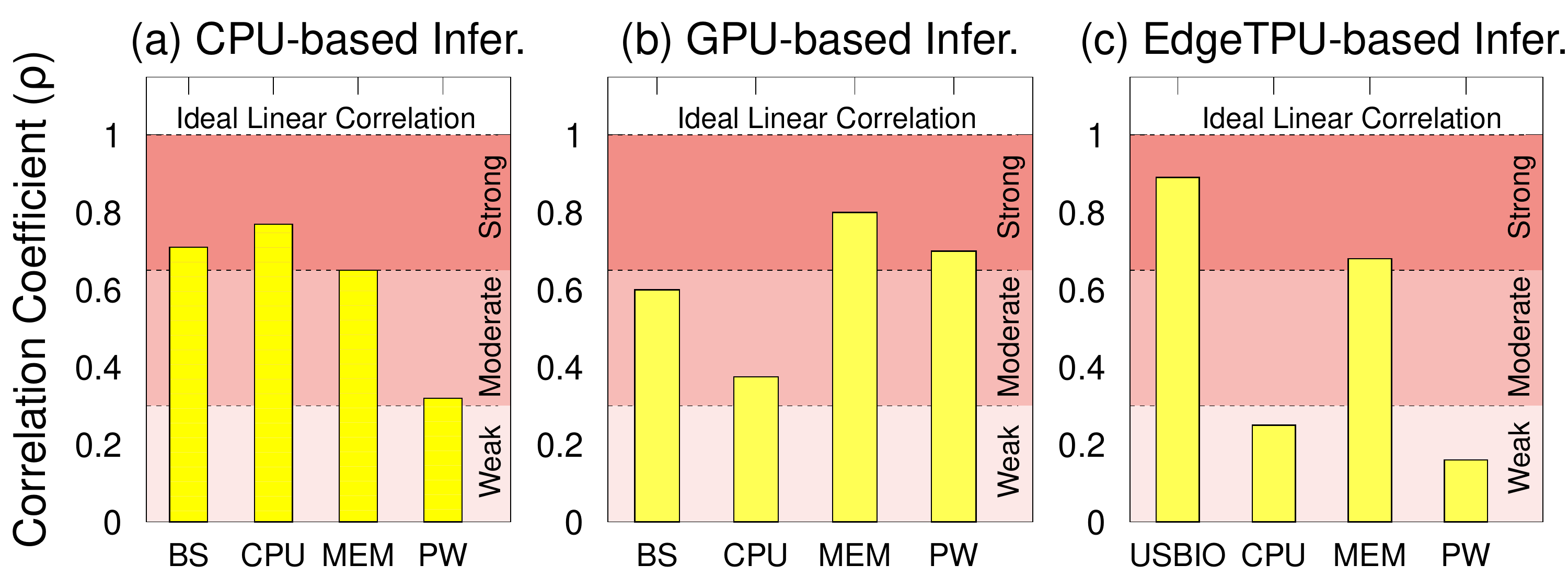}
	\caption{Correlated factors that change the inference throughput. (BS: Batch Size, CPU: CPU usage, MEM: memory usage, PW: Power consumption, USBIO: USB IO bandwidth usage) }
	\label{fig:corr_factors}
\end{figure}

Fig.~\ref{fig:corr_factors} shows the correlated factors for the DL inference throughput when using CPU, GPU, and \TPU. 
For the CPU-based inferences (e.g., \RPI, \ODN), the CPU, batch size, and memory were strongly correlated with the inference throughput results. 
CPU resources were mainly used to perform the DL tasks, and memory resources were used to load and store the DL models. 
The inference tasks with larger batch sizes naturally increased the input data for processing so that an increase in the batch sizes could improve the throughput until the limit of device resources. 

\begin{figure}[t]
    \centering
	\includegraphics[width=1\columnwidth]{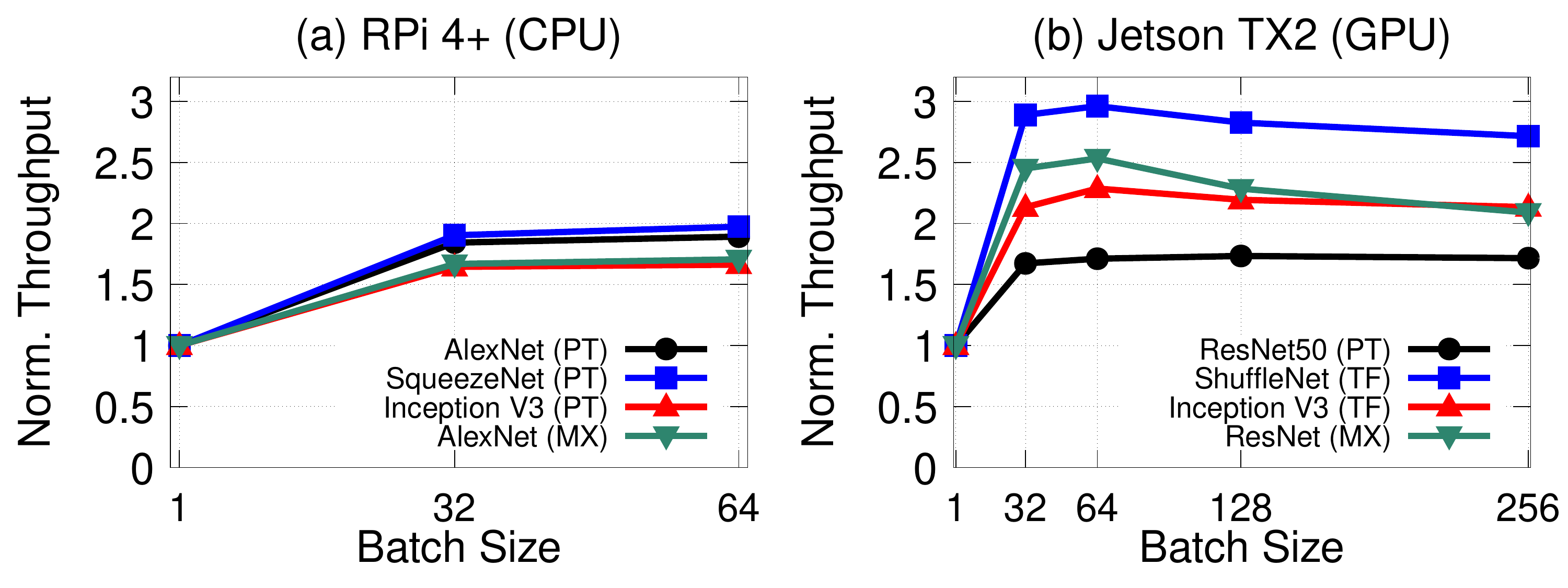}
	\caption{Throughput changes with different batch sizes.}
	\label{fig:batch_impact}
\end{figure}

For the GPU-based inference tasks (e.g., \Nano and \TX), memory, power, and batch sizes were positively correlated with the DL inference throughput.
Specifically, the power consumption showed a strong correlation with the throughput as the GPU module in edge devices consumed more power than typical CPUs in edge devices. 
And, CPU showed a relatively weaker correlation with the throughput as CPU was only used for managing the device and processes co-running (non-DL) applications rather than performing the DL tasks. 
Because the batch size showed a strong correlation for both the CPU and GPU-based inference tasks, we report the impact of batch size changes in Fig.~\ref{fig:batch_impact}. 
As shown in the results, the batch sizes changed the DL inference throughput significantly. 
In general, a larger batch size appeared to result in increased throughput; however, an interesting observation is that using larger batch sizes did not always increase the DL inference throughput.
This suggests that employing the right (or optimal) size of the input batch will be critical for improving the DL inference throughput on edge devices.

In the \TPU-based inferences cases, the USB bandwidth (between a host edge device and the \USBA) and memory usage on host edge devices strongly correlated with the inference throughput. 
Both memory and USB IO were closely related to each other for executing DL models on the \USBA. 
Because \USBA does not have main memory (RAM)\footnote{\USBA has only 8MB of cache memory (SRAM).}, it relies on the host device's memory system to store models and uses context switching to swap models/parameters between the host device's RAM and \TPU to perform DL inference tasks. Therefore, low USB IO bandwidth between the host device and \USBA limits data rates for switching models and parameters so that the throughput can decrease. 

\begin{figure}[t]
    \centering
	\includegraphics[width=1\columnwidth]{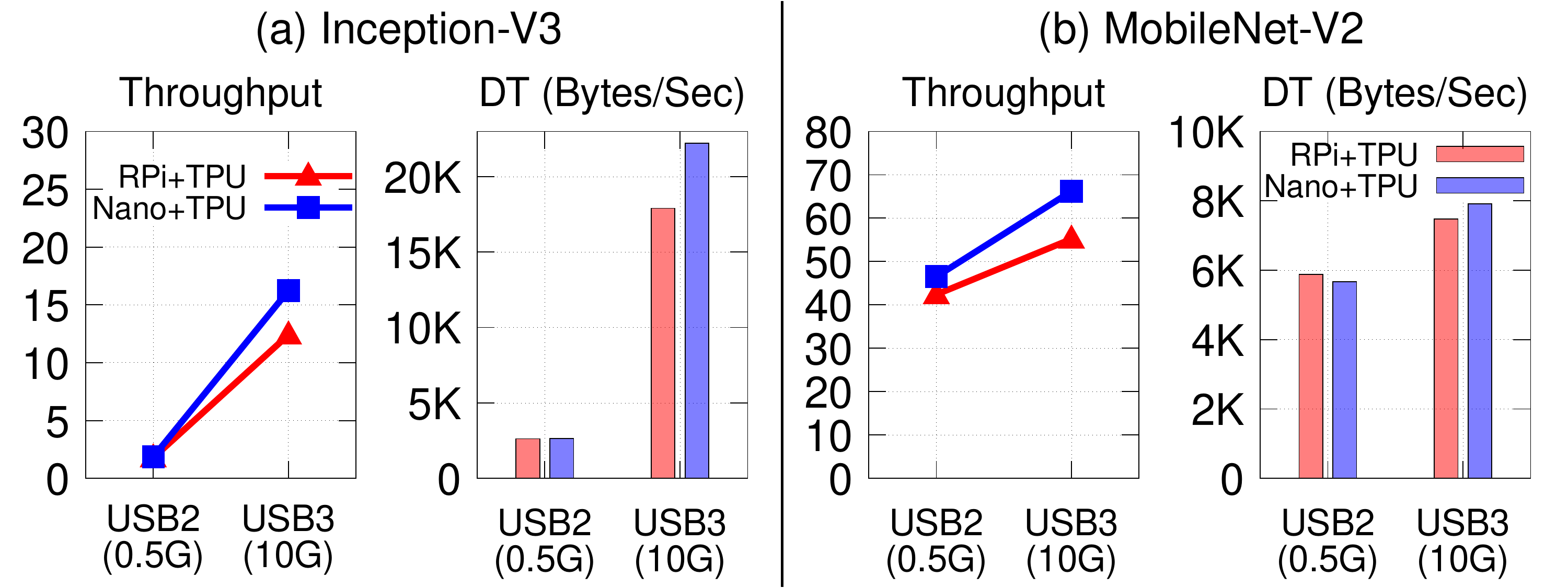}
	\caption{Difference in DL inference throughput and data transfer with USB 2.0 and 3.0 interfaces. (DT: Data Transfer Amount)}
	\label{fig:usb_impact_merge}
\end{figure}

To further investigate the impact of the USB IO bandwidth, we measured the DL inference throughput changes from \USBA by connecting it with two edge devices (\RPI and \Nano). 
Also, to observe the throughput changes with different bandwidth, we used two USB interface types.
i.e., USB 2.0 with up to 0.5GB of bandwidth, USB 3.0 with up to 10GB of bandwidth.
As shown in Fig.~\ref{fig:usb_impact_merge}, the results confirm that USB's IO bandwidth could considerably change the DL inference throughput of \TPUs. 
With larger IO bandwidth supported by USB 3.0, \RPI achieved 1.3$\times$ (\mobtwo) and 7$\times$ (\incept) higher throughput than the inference with USB 2.0.
\Nano also showed 1.4$\times$ (\mobtwo) and 8.7$\times$ (\incept) higher throughput than \USBA with USB 2.0.
Larger USB IO bandwidth facilitated the switching of model parameters and input data between the host device and \USBA so that it significantly improved the overall DL inference throughput.

\vspace{0.75mm}
\noindent
{\bf Summary.} 
This section characterized the performance and behaviors of edge devices and \TPU accelerators with AI single-tenancy, focused on the inference throughput.
We found several factors that changed the DL inference throughout as well as identified correlated resources for the throughput changes. 
In the next section, these results will be used as baselines for evaluating and characterizing the AI multi-tenancy on edge devices and \TPUs.

\section{Evaluation of AI Multi-Tenancy on Edge Devices and Accelerators}\label{sec:multi-tenancy}
This section evaluates and characterizes two techniques for enabling AI multi-tenancy on edge devices and \TPUs.

\subsection{AI Multi-Tenancy with Concurrent Model Executions}\label{subsec:cme}
Concurrent model executions (CMEs) leverage the idea of parallel processing and enable AI multi-tenancy by simultaneously executing multiple DL inference tasks on edge devices' resources. 
e.g., deploying and executing multiple DL models on either GPU or \TPUs.
CME can provide two potential benefits to edge devices and \TPUs; 1) improving DL inference throughput and 2) allowing to run multiple (often different) DL services (e.g., inference tasks).
Therefore, it is important to correctly identify the upper bound of throughput improvement and the concurrency level (the number of co-running DL models) on the devices' resources by CME.
Moreover, the maximum concurrency level may not provide the highest throughput, so it is also important to determine the concurrency level that results in the highest throughput.
Therefore, we performed an empirical evaluation of CME with DL models 
to answer the following questions;
\begin{enumerate}
    \item What is the maximum DL inference throughput of the edge devices and \TPUs with CME?
    \item What is the maximum concurrency level on the edge devices and \TPUs with CME?
    \item What is the concurrency level on edge devices and \TPUs to maximize DL inference throughput?
\end{enumerate}

In this evaluation, we used three DL models (e.g., \incept, \mobone, \mobtwo) for evaluating CME because all these models could be executed on three resource types in edge devices and \TPU accelerators.
Among all DL frameworks, we excluded TF from this CME evaluation since TF is not thread-safe.
Specifically, {\tt kerascv}~\cite{kerascv} and {\tt tf.Graph}~\cite{tf.Graph} libraries did not fully support concurrent executions.
Regarding the throughput calculation with CME, we changed equation-(\ref{eq:throughput_single}), and the number of inferences in the equation was calculated by ``concurrency level'' $\times$ ``batch size'' $\times$ ``the number of batches.''

\begin{figure*}[t]
\centering
\begin{minipage}[b]{1.0\columnwidth}
	\includegraphics[width=0.98\columnwidth]{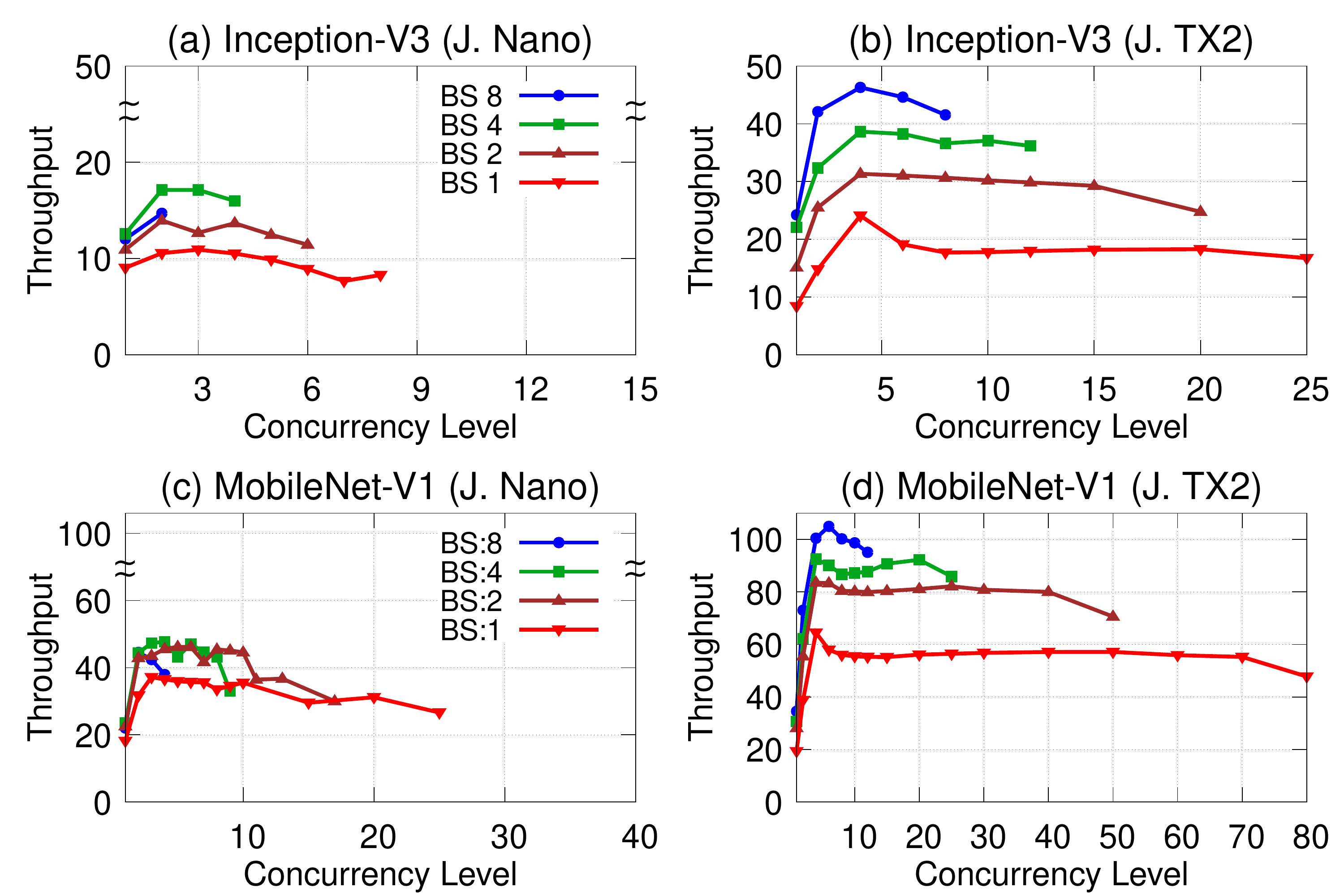}
	\caption{CME measurement results of throughput and concurrency level on GPUs with PyTorch (BS: Batch Size)}
	\label{fig:gpu_cc_pt}
\end{minipage}
\smallskip
\begin{minipage}[b]{1.0\columnwidth}
	\includegraphics[width=0.98\columnwidth]{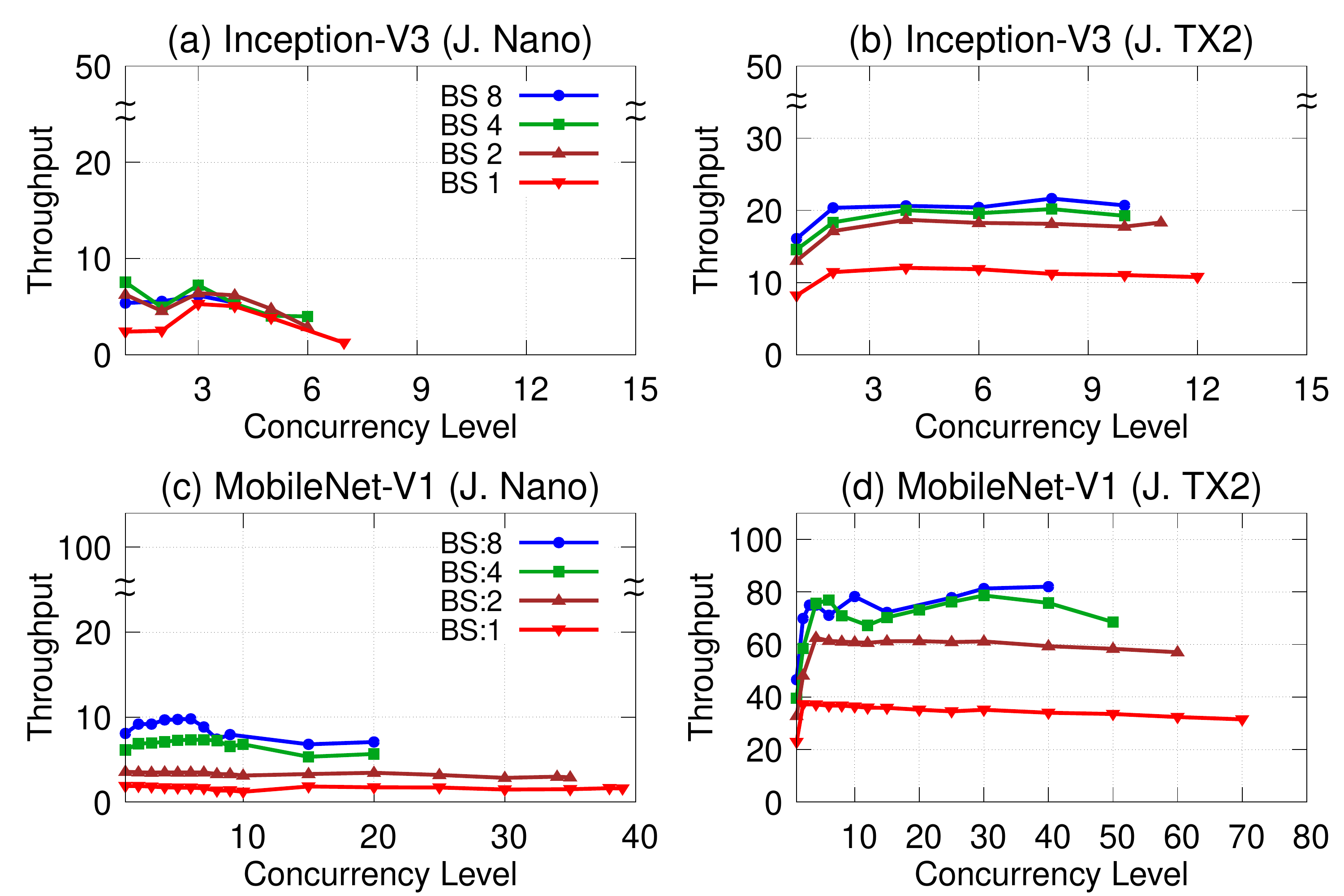}
	\caption{CME measurement results of throughput and concurrency level on GPUs with MXNet (BS: Batch Size)}
	\label{fig:gpu_cc_mx}
\end{minipage}
\vspace{-1.0em}
\end{figure*}

\vspace{1mm}
\noindent
\textbf{Evaluation Steps for CME.}
We began the CME evaluation by deploying and executing a single DL model on edge devices and \TPU. 
We then gradually increased the number of co-running DL models (``{\em concurrency level}'') on the devices and \TPUs to measure the changes in the DL inference throughput and resource usage patterns. 
This experiment was continued to increase the concurrency level until the benchmarker failed to run. 
The concurrency level obtained from the last successful execution was considered as the maximum concurrency level supported by the edge devices and \TPUs. 
In this measurement, we only report the results with leveraging CME on GPUs (\Nano and \TX) and \TPUs (\DevB and \USBA), and we omit the measurement results from CPU resources.
This is because, while we could find some benefits of CME on CPUs, e.g., six concurrent models could be executed on CPUs of \RPI and \ODN, the throughput benefits were marginal, and the measured throughput results were exceptionally lower than the results with CME on either GPUs or \TPUs.

\vspace{1mm}
\noindent
{\bf CME Evaluation Results on GPUs.}
Fig.~\ref{fig:gpu_cc_pt} and \ref{fig:gpu_cc_mx} show DL inference throughput changes with different concurrency levels on GPUs in \Nano and \TX. 
Please note that, in both graphs, we omit the results from \mobtwo due to the page limit, and the results were similar to the results with \mobone.

\begin{figure}[t]
    \centering
	\includegraphics[width=1\columnwidth]{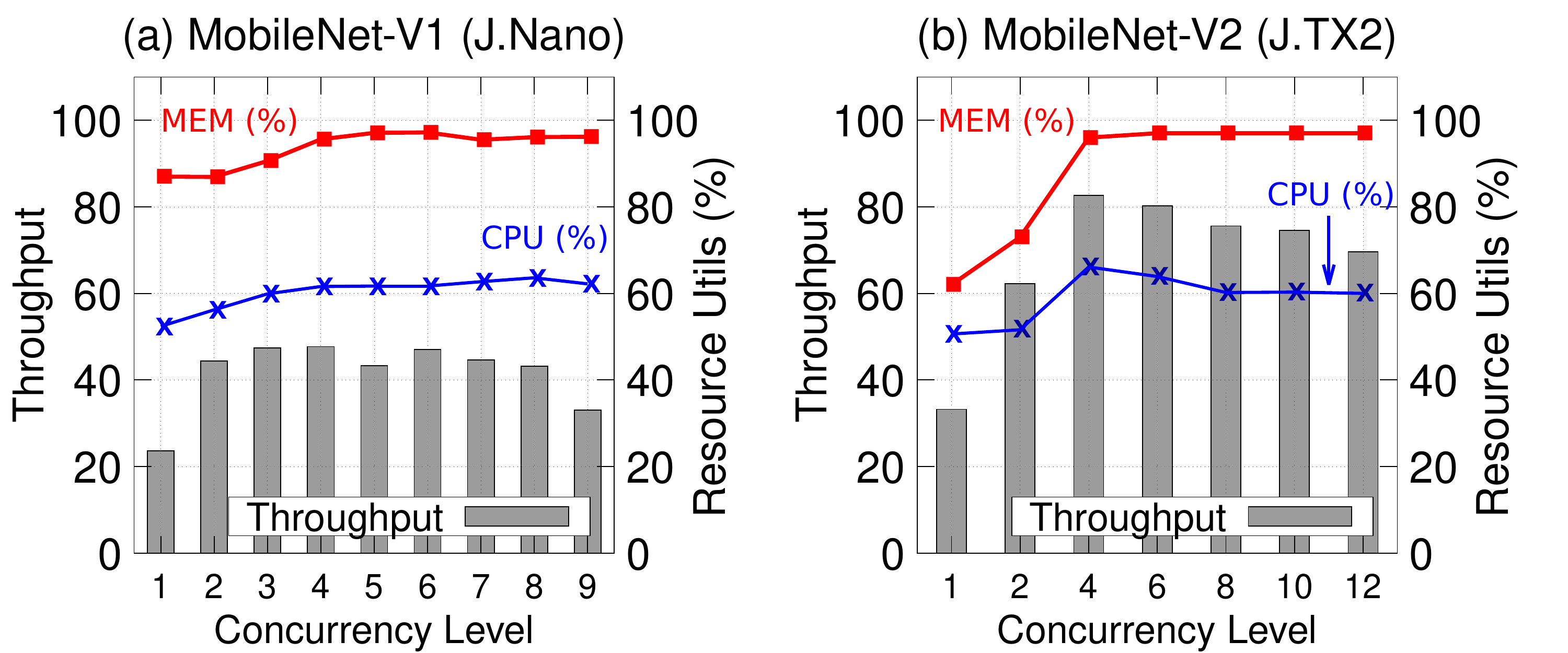}
	\caption{Resource utilization and throughput changes with CME (PyTorch). \Nano uses a batch size of 4, and \TX employs a batch size of 8.}
	\label{fig:GPU-PT-Res}
 	\vspace{-0.5em}
\end{figure}

When enabling CME on GPU using PyTorch (shown in Fig.~\ref{fig:gpu_cc_pt}), the maximum concurrency level and throughput varied with different batch sizes. 
DL inference with a batch size of 1 provided the maximum concurrency level. 
We observed that \Nano could run 8 (\incept) to 25 (\mobone) models concurrently on GPU, and \TX was able to deploy 25 (\incept) to 80 (\mobone) models on its GPU simultaneously.
Using larger batch sizes (e.g., batch size of 4 for \Nano, batch size of 8 for \TX), the multi-tenancy enabled by CME significantly improved the DL throughput against the single-tenancy cases. 
In particular, with CME, \Nano showed 1.3$\times$ to 1.9$\times$ improved throughput, and \TX showed 1.7$\times$ to 2.7$\times$ higher throughput against the single-tenancy cases.
Our further analysis revealed that memory resource was the critical factor to determine the maximum throughput when enabling CME. 
As Fig.~\ref{fig:GPU-PT-Res} shows, the maximum throughput was highly correlated with memory utilization. 
Both \Nano and \TX showed that the maximum throughput was reached when the memory resource was saturated. After reaching the maximum throughput, the throughput was either decreased or stabilized with high memory utilization.
It is worth noting that the high correlation between memory utilization and throughput increase was consistent with our observation reported in Fig.~\ref{fig:corr_factors} in the previous section.

However, the CME evaluation with MXNet showed different results from the previous measurements with PyTorch. 
Both \Nano and \TX had lower throughput improvement. 
In particular, \Nano showed considerably low performance, and on average, \Nano with MXNet had even 13\% lower throughput than single-tenancy cases. 
This low throughput was because \Nano's experiments were performed by disabling MXNet's {\tt cuDNN auto-tune}~\cite{MXNet-EnvVar} parameter so that the framework used sub-optimal convolution layers for {\tt cuDNN}. 
Enabling or disabling {\tt auto-tune} option can significantly impact DL inference throughput because, if this option is enabled, MXNet first runs a performance test to seek the best convolutional algorithm, and the selected algorithm is used for further inference tasks.
However, this performance test requires significant consumption of resources on edge devices. Unfortunately, due to the smaller memory size (4GB), \Nano could not complete this performance test due to the frequent out-of-memory errors.

For \TX, while the throughput benefits using CME were smaller than the throughput with PyTorch, it showed 1.12$\times$ to 1.5$\times$ higher throughput compared to the single-tenancy cases.
Regarding the concurrency level supported by CME with MXNet, \Nano successfully ran 6 (\incept), 39 (\mobone), and 45 (\mobtwo) concurrent models, and \TX could run 12 (\incept) and 70 (\mobone and \mobtwo) models with a batch size of 1.

\begin{figure}[t]
    \centering
	\includegraphics[width=1\columnwidth]{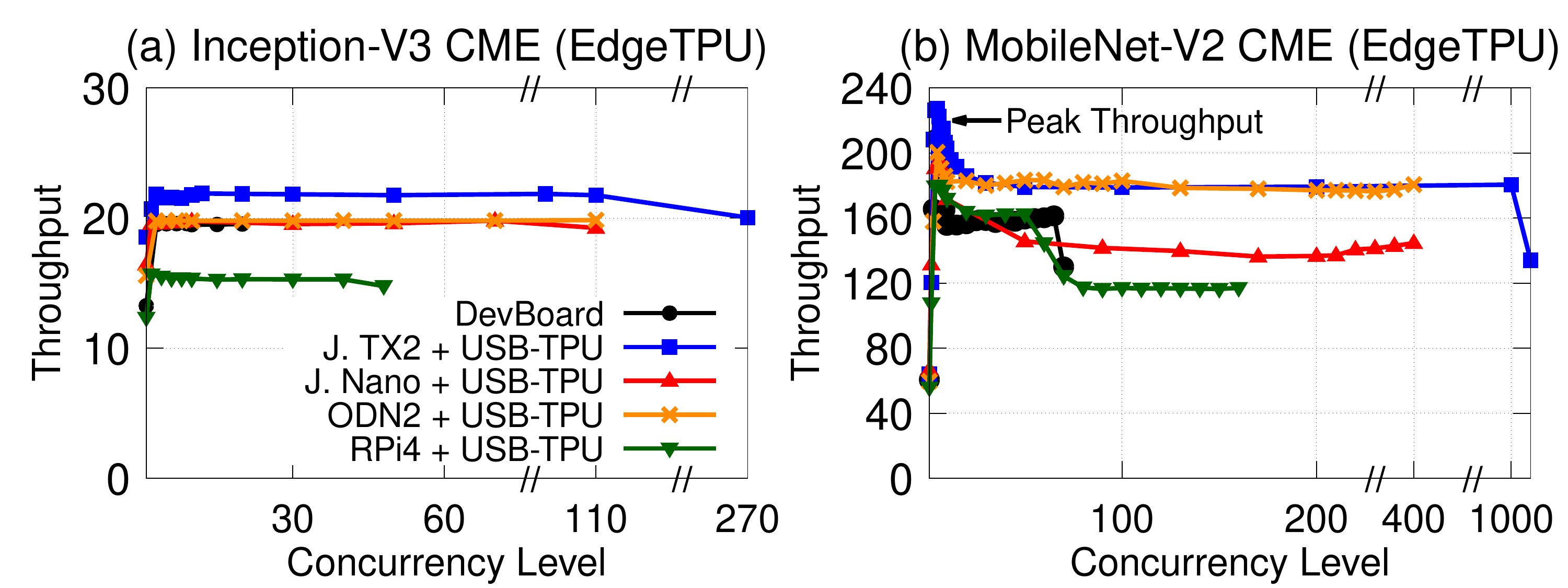}
	\caption{DL inference throughput variations by enabling CME on \TPUs}
	\label{fig:TPU_concurrency}
	\vspace{-0.5em}
\end{figure}

\vspace{1mm}
\noindent
{\bf CME Evaluation Results on \TPUs.} 
Fig.~\ref{fig:TPU_concurrency} reports DL inference throughput variations with different concurrency levels on \TPUs (both \DevB and \USBA).
\USBA's throughput and concurrency level results were measured using four edge devices. 
Similar to the previous results on GPUs, CME on \TPUs could also increase throughput over the single-tenancy cases.
For \incept (Fig.~\ref{fig:TPU_concurrency}(a)), \DevB had 1.3$\times$ higher throughput, and \USBAs showed 1.25$\times$ improved throughput over single-tenancy cases. For both \mobone and \mobtwo (Fig.~\ref{fig:TPU_concurrency}(b)), \TPUs showed 3.3$\times$ higher throughput over the single-tenancy cases.
Please note that we omit the throughput results with \mobone because the results are similar to \mobtwo.

In this evaluation, we found two interesting observations about the throughput improvement.
One is that CME's throughput increase with \incept (1.3$\times$) was much smaller than \mobboth (3.3$\times$).
The other is that \mobboth reached the maximum throughput with lower concurrency levels, and the throughput is decreased and stabilized with higher concurrency levels. 
Our further analysis revealed that the above two issues were related to the model size and \TPU's 8MB of SRAM used to cache the DL model's parameters.
In particular, a smaller throughput increase with \incept was because 25MB of \incept size could not be fully loaded in the \TPUs' cache (SRAM), and thus, model parameter swapping operations between the \TPU's cache and the edge devices' memory were continuously being performed.
Therefore, the increased concurrency level did not increase the inference throughput due to the high overhead in the model parameter swaps.
On the other hand, if the model size was small, e.g., 4MB of \mobtwo, the model could be fully loaded in \TPUs' cache and did not require frequent operations of model parameter swapping, hence low USB IO overhead and possibly significant throughput increases. 

Regarding the second observation found in Fig.~\ref{fig:TPU_concurrency}(b), the \TPU cache could load even multiple smaller models simultaneously. While \TPU could leverage only one model at a time, other loaded models were able to obtain data from the host device's memory, hence minimizing the delay when switching models in \TPUs. 
On the other hand, if the concurrency level was high, frequent model swaps needed to be frequently performed in \TPU's cache, resulting in increased data transfer between \TPU and the host edge device's memory. Therefore, USB IO was quickly saturated, and throughput could be degraded. 
This is why both \mobone and \mobtwo reached the maximum throughput with a low concurrency level, and throughput could be decreased and stabilized with higher concurrency levels.
This analysis suggests that, when using CME on \TPU, model size and concurrency level should be carefully determined to increase the throughput. 
Moreover, model compression techniques~\cite{DLwithEdge-PIEEE-2019}, e.g., quantization and parameter pruning, should be considered for optimizing model size for \TPUs.

\begin{figure}[t]
    \centering
	\includegraphics[width=1\columnwidth]{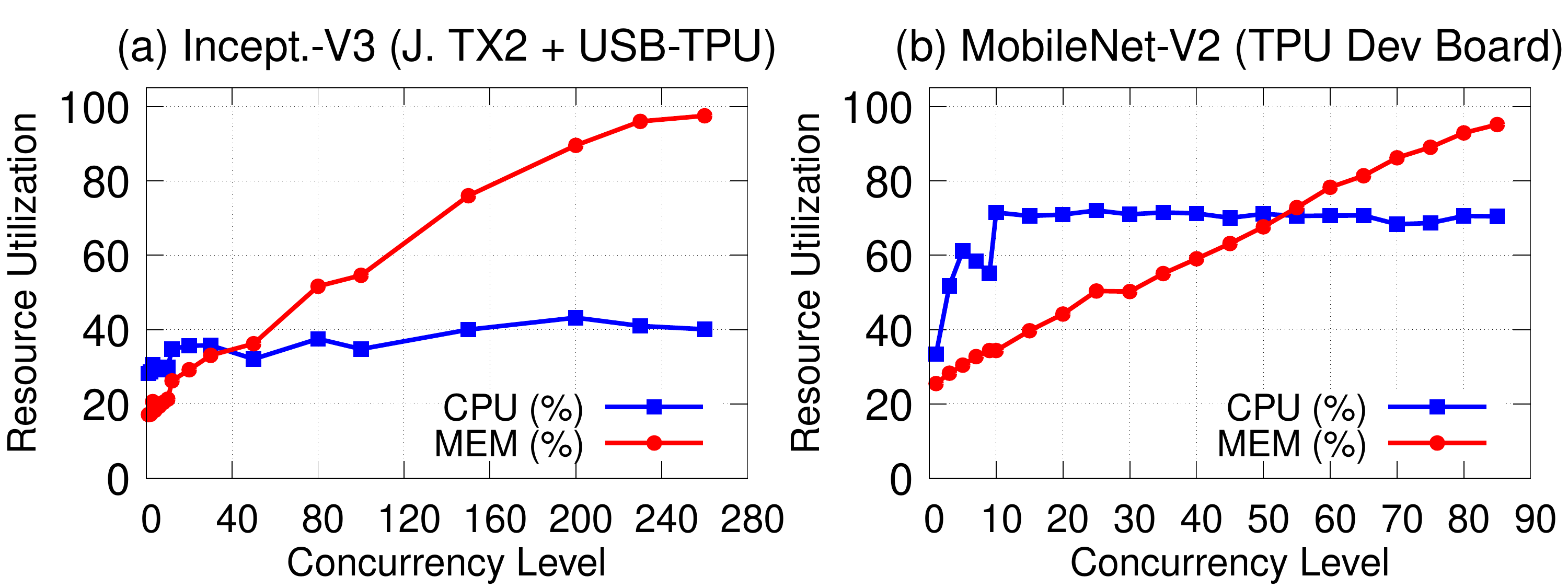}
	\caption{Resource utilization changes with increased concurrency level (\TPUs)}
	\label{fig:TPU-CC-Res}
	\vspace{-0.5em}
\end{figure}

The three models reported much higher concurrency levels on \TPUs than the concurrency level on GPUs.
\DevB supported the concurrency level of 20 for \incept and the concurrency level of 80 -- 85 for both \mobboth models.
Furthermore, \USBAs reached various maximum concurrency levels. Specifically, \USBAs' concurrency levels varied considerably across different host edge devices. 
For example, for \incept, the maximum concurrency level from \USBA with \RPI was 48, but when it used \TX as the host device, the maximum concurrency level reached 270. For \mobtwo, the maximum concurrency level from \USBA with \RPI was 160, but it could be 1100 when leveraging \TX as the host edge device. 
Regarding the varying concurrency levels, our further analysis revealed that the maximum concurrency levels supported by \USBAs had a high correlation with the size and utilization of memory resources in the host edge devices. 
Fig.~\ref{fig:TPU-CC-Res} shows resource utilization changes with different concurrency levels measured from \USBA with \TX and \DevB. 
The results show that memory utilization increased as the concurrency level went up. 
The maximum concurrency level was determined when the memory utilization reached close to 100\%, indicating that memory size and bandwidth often limit the supported concurrency level DL models when enabling CME on \USBA.

\subsection{AI Multi-Tenancy with Dynamic Model Placements}\label{subsec:dmp}
\begin{figure*}[t]
    \centering
	\includegraphics[width=1\textwidth]{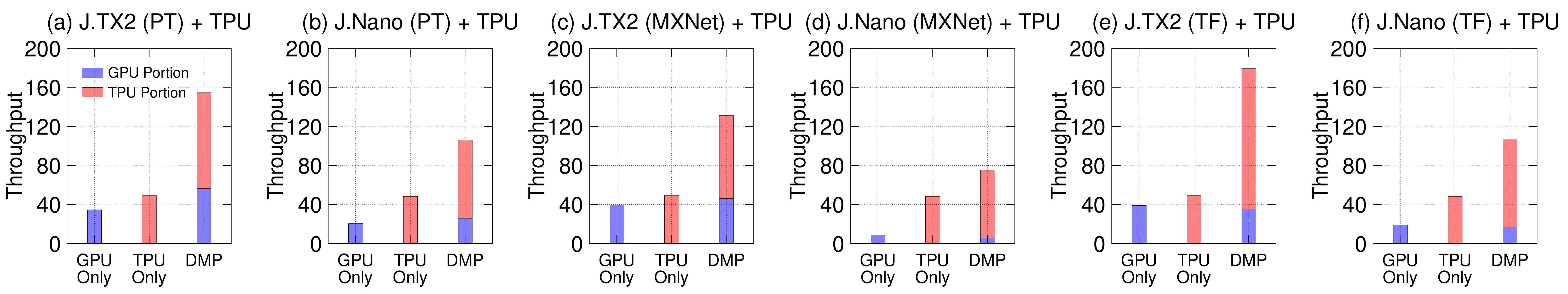}
	\caption{Comparison of DL inference throughput between DMP and single-tenancy.}
	\label{fig:DMP_comp_single}
\end{figure*}

\begin{figure*}[t]
    \centering
	\includegraphics[width=1\textwidth]{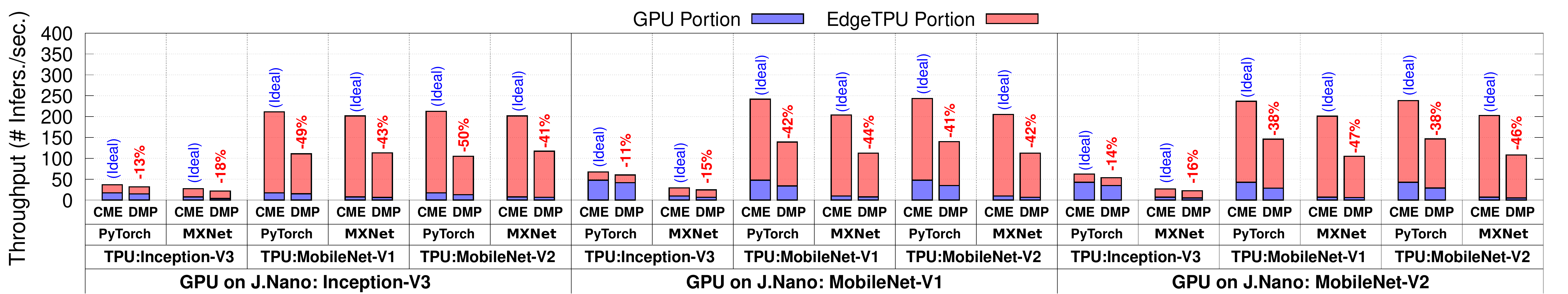}
	\caption{\Nano's DL inference throughput comparison between (ideal) results from CME and DMP. The (ideal) results from CME are calculated by the sum of CME throughput on GPU and CME throughput on \TPU, which were measured separately.}
	\label{fig:DMP_comp_cme_jnano}
\end{figure*}

\begin{figure*}[t]
    \centering
	\includegraphics[width=1\textwidth]{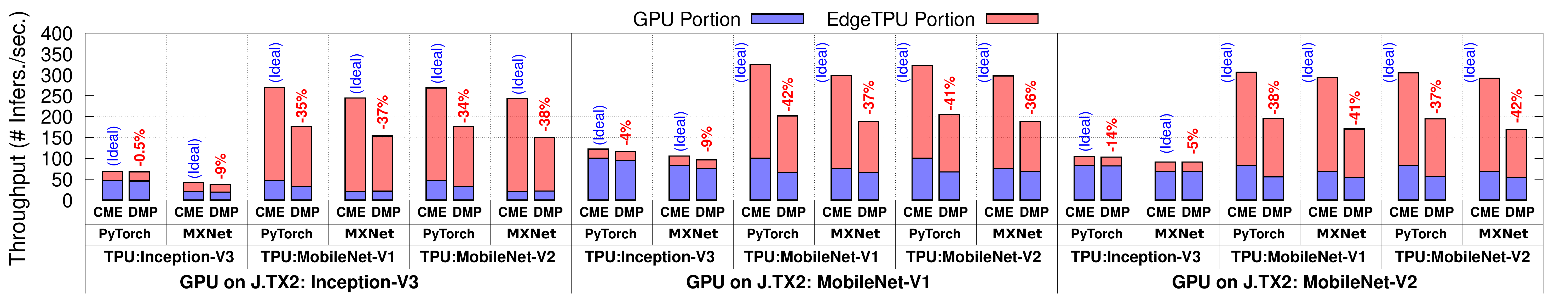}
	\caption{\TX's DL inference throughput comparison between (ideal) results from CME and DMP. The (ideal) results from CME are calculated by the sum of separately measured CME throughput on GPU and \TPU.}
	\label{fig:DMP_comp_cme_jtx2}
\end{figure*}

This section characterizes and evaluates the dynamic model placement (DMP) technique for AI multi-tenancy on edge devices and \TPUs. 
DMP allows running multiple DL models simultaneously by placing DL models on an edge device's resource (CPU and/or GPU) and other DL models on \TPUs.
Because \USBA can be connected to edge devices via USB interfaces, the potential benefits from DMP can be improved DL inference throughput using heterogeneous resources in both edge devices and \USBA as well as high resource utilization of both resources.
However, DL inference tasks from both on-board edge resources and \USBA are managed by the host edge devices so that there can be a performance penalty from resource contention.
Therefore, in this evaluation, we focus on seeking answers to the following research questions;
\begin{enumerate}
    \item What are the performance benefits (e.g., DL inference throughput) from DMP on heterogeneous resources?
    \item What are the actual performance penalties of using DMP, compared explicitly to CME for AI multi-tenancy?
\end{enumerate}

Similar to the CME evaluations, we used three DL models (\incept, \mobone, and \mobtwo) because these models could perform inference tasks on all resource types in edge devices and \TPUs. 
We also changed the equation-(\ref{eq:throughput_single}) to correctly calculate the throughput with DMP. Specifically, the number of inferences for DMP was calculated by the sum of the inference numbers from edge resources (CPU or GPU) and the inference numbers from \USBA.

We initially used four edge devices (\RPI, \ODN, \Nano, \TX) connected with a \USBA and deployed DL models on both edge resources and the \USBA. 
However, we omit the evaluation results of \RPI and \ODN because we could not observe the benefits of using DMP on such devices. 
Specifically, CPUs on \RPI and \ODN were quickly saturated by both CPU-based and \TPU-based DL inference tasks, and the overall inference throughput results with DMP on \RPI and \ODN could be even lower (about 10\%) than \TPU-only inference throughput.

We evaluated DMP with three DL frameworks for GPUs and TFLite for \TPU.
We enabled CME when the models are running on PyTorch (GPU), MXNet (GPU), and TFLite (\TPU), and we used single-tenancy with TF on GPU.

\vspace{1mm}
\noindent
\textbf{DMP Evaluation Results.} 
Fig.~\ref{fig:DMP_comp_single} shows DMP's DL inference throughput improvement against the single-tenancy cases. 
Both \Nano and \TX showed significantly increased throughput compared to the single-tenancy GPU or \TPU-based inferences.
In particular, \Nano had 6.2$\times$ improved throughput over the single-tenancy on GPU and 2$\times$ increased throughput over the single-tenancy on \TPU.
\TX also showed throughput improvement by 3.8$\times$ (for GPU) and 3.1$\times$ (for \TPU).
However, this improved throughput can be in part due to leveraging both CME and DMP. 
We also compare the DL inference throughput between ideal throughput upper bound based on CME results (reported in Section~\ref{subsec:cme}) and DMP.
Please note that the ideal throughput upper bound is calculated by accumulating GPU throughput with CME and \TPU throughput with CME measured separately. 

Fig.~\ref{fig:DMP_comp_cme_jnano} and \ref{fig:DMP_comp_cme_jtx2} report the throughput comparison between (ideal) CME results and DMP.
The results contain the results measured from \Nano and \TX when using PyTorch/MXNet (for GPU) and TFLite (for \TPU). 
As shown in the figures, while the differences between the ideal throughput and DMP's throughput varied with DL models and DL frameworks, \Nano with DMP (Fig.~\ref{fig:DMP_comp_cme_jnano}) and \TX with DMP (Fig.~\ref{fig:DMP_comp_cme_jtx2}) showed 34.6\% and 25.3\% lower DL inference throughput than the ideal throughput with CME (on both GPU and \TPU).
Such differences were mainly due to the resource contention and resource limits in the edge devices.

\begin{figure}[t]
    \centering
	\includegraphics[width=1\columnwidth]{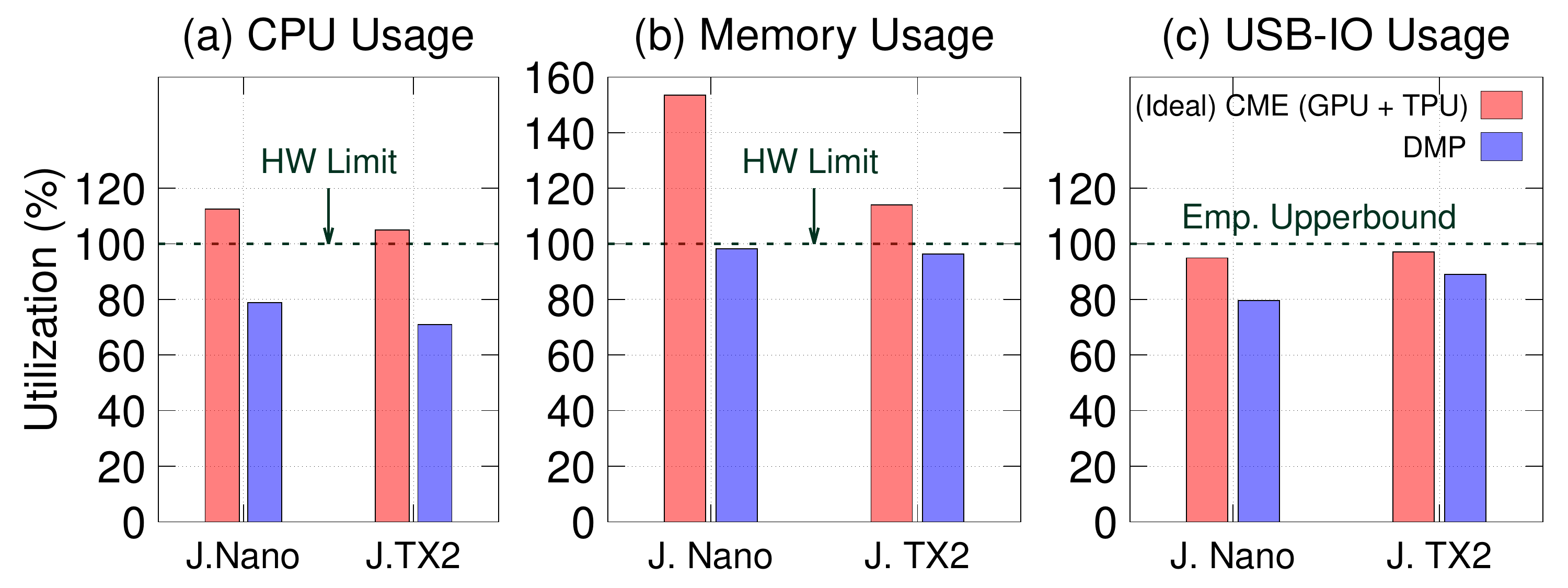}
	\caption{Resource usage comparison between (ideal) sum of CME on GPU/\TPU and DMP.}
	\label{fig:DMP-Res}
\end{figure}

To understand the gap between the DMP's throughput and ideal throughput, we performed further analysis on resource consumption.
Fig.~\ref{fig:DMP-Res} shows the resource utilization (CPU, memory, USB IO) between the ideal sum of CME on GPU/\TPU and DMP. 
As shown in the figure, the ideal throughput often could not be achievable with current HW specifications. 
Specifically, CPU and memory utilization should exceed the HW limits of the edge devices (more than 100\%) to reach such high throughput. 
Moreover, similar to the CME analysis, memory was identified as a critical resource when enabling DMP. 
Specifically, we observed that memory utilization reached 100\% with DMP, but CPU utilization did not reach 100\%. 
Based on this observation, the DL inference throughput, when the memory resource is saturated, can be the empirical performance upper bound when enabling DMP.
We also observed that resource contention could impact the DL inference throughput because the shared resources, such as memory and CPU, were needed to manage multiple DL models running on different resources. 
The decreased USB IO utilization (about 8\% to 15\%) with DMP (Fig.~\ref{fig:DMP-Res}(c)) was because of such resource contentions, and the reduced USB IO utilization could decrease DL inference throughput from \TPU in \USBA.

\section{AI Multi-Tenancy Summary and Discussion}\label{sec:discussion}
This section describes a summary of our evaluation results with CME and DMP and discusses important findings regarding AI multi-tenancy on edge devices.

When enabling CME on GPUs for AI multi-tenancy, we observed significant DL throughput improvement over single-tenancy cases.
In particular, \Nano and \TX showed 1.3$\times$ to 2.7$\times$ improved throughput, and both devices support the concurrency levels of 8 (\incept on \Nano) to 80 (\mobone on \TX). 
However, we found that high concurrency levels did not necessarily result in improving DL inference throughput.
Moreover, both concurrency level and batch sizes can significantly change the DL inference throughput, so both configurations should be carefully determined and optimized when applying CME on GPUs.
We also observed that the maximum throughput with CME on GPU could be determined when the memory utilization reached 100\%. 
Finally, the throughput benefits of CME on GPUs can vary across different DL frameworks and edge devices. 
Sometimes, the throughput benefit can be limited, especially when applying CME on GPU with MXNet on edge devices with small memory sizes like \Nano.

When enabling CME on \TPUs, our evaluation confirmed considerable performance benefits, and high concurrency levels could be achieved.
However, similar to CME on GPUs, high concurrency levels did not necessarily result in the maximum DL inference throughput.
We observed that DL inference throughput could be affected by the DL model size and the small cache size in \TPUs. 
Specifically, if the DL model size could not fit in the small cache (8MB) of \TPUs, the benefits of using CME could be limited. 
Moreover, lower concurrency levels often resulted in a maximum performance gain when using small DL models (e.g., \mobboth).
This observation strongly suggests that the techniques for minimizing model size (e.g., quantization and model compression) will be critical for the throughput improvement.
Moreover, we observed that the maximum concurrency level could be determined when the memory utilization reaches near 100\%. 
This observation indicates that host edge devices' memory capacity can be critical for increasing the concurrency level when using \USBA.

When enabling DMP for AI multi-tenancy, our evaluation confirmed that DMP considerably improved DL inference throughput over single-tenancy cases by leveraging GPU and \TPU resources simultaneously.
We also compared the DMP's throughput against the ideal inference throughput, which was the sum of the maximum throughputs when applying CME on both GPU and \TPU.
While the DMP's maximum throughput showed 25\% to 35\% lower throughput than the ideal inference throughput, such differences were mainly due to the HW limitations.
We also observed that memory could be the critical resource factor when enabling DMP, and memory resources were saturated when reaching the maximum throughput.
We found the impact of resource contention with DMP. In particular, USB IO bandwidth usage was decreased when applying DMP, and the reduced usage resulted in lower throughput from \TPU.

\section{Related Work}\label{sec:related}
Several studies have been conducted to quantify the performance of various edge devices for DL/ML inference tasks~\cite{pCAMP-HotEdge-2018, EmBench-CoRR19, ResCharct-AUChallengeIoT19, PerfUSBAccelerator-2020, CharacterizingDNNonEdge-IIWSC2019, AIEdge-CoRR-2020}. 
However, most studies have focused on characterizing performance (e.g., latency and throughput) and efficiency (e.g., energy consumption) of edge devices and AI accelerators with single DL tasks.

pCamp~\cite{pCAMP-HotEdge-2018} evaluated ML packages and frameworks' performance when executing image classification tasks on edge platforms, including \TX, \RPI, and Nexus 6p. 
This work reported latency (including model loading time), memory usage, and energy consumption from different ML packages. 
Hadidi et al.~\cite{CharacterizingDNNonEdge-IIWSC2019} have characterized various edge devices and AI accelerators (e.g., \TPUs) with DL inference tasks. The authors analyzed the impact of DL frameworks and SW stacks as well as measured energy consumption and temperature when performing DL inference tasks.
Moreover, several studies~\cite{ResCharct-AUChallengeIoT19, EmBench-CoRR19, CrossPlatformBench-MLSYS20} have focused on characterizing the performance of DL inference tasks on different HW architectures and resource models (CPU, GPU, \TPU). 
EmBench~\cite{EmBench-CoRR19} has performed  a per-layer analysis of DL inference tasks to identify performance bottlenecks. Libutti et al.~\cite{PerfUSBAccelerator-2020} conducted performance evaluations of DL inference tasks with portable, USB-based edge accelerators, including Coral USB Accelerator and Intel Neural Compute Stick~\cite{Neural-Sticks}.

More recently, Liang et al.~\cite{AIEdge-CoRR-2020} have conducted an experimental study to evaluate model splitting and compression techniques on edge devices and accelerators.
Network latency, bandwidth usage, and resource utilization with various configurations were also reported when applying model splitting and compression to cloud-edge co-inference use-cases.
Additionally, the authors have evaluated the concurrency model executions for multi-tenancy use cases. 
However, the concurrency evaluation is narrowly performed with only one model having a single batch size. Moreover, in addition to evaluating the CME strategy, our work also evaluated and characterized the DMP strategy for AI multi-tenancy that leverages heterogeneous resources in edge and \TPU.

\section{Conclusion}\label{sec:conclusion}
In this work, we evaluated the performance of two AI multi-tenancy techniques for edge computing; CME (Concurrent Model Executions) and DMP (Dynamic Model Placements).
These two techniques were evaluated on various edge devices and \TPU accelerators with widely used DL frameworks and models for image classification tasks.
Our evaluation confirmed that both AI multi-tenancy techniques could significantly improve the DL inference throughput. 
We empirically identified the maximum concurrency level of DL models supported by various edge devices and \TPU accelerators for the CME technique. 
Moreover, we also investigated resource factors impacting the concurrency level with CME.
We also identified the benefits and limitations of DMP on heterogeneous resources. Specifically, memory resources can be the critical resource factor determining the maximum DL inference throughput with DMP. 
Additionally, we observed that the USB IO bandwidth usage could be decreased due to the resource contention from DMP.

Both techniques open up new opportunities for AI multi-tenancy, including flexible and high-performance DL service deployment. 
Furthermore, further research efforts need to be followed to maximize the benefit of AI multi-tenancy, such as safe placement of multiple DL models to minimize the resource contention and a better isolation mechanism for dynamic control of DL inference throughputs.

\section*{Acknowledgment}\label{sec:ack}
This research was in part supported by NSF grant SES-1637277 and USDA grant 2021-67019-34342. Any opinions, findings, conclusions, or recommendations expressed in this publication are those of the authors and do not necessarily reflect the view of NSF and USDA.

\bibliographystyle{unsrt}
\bibliography{bibfiles/reference.bib}

\end{document}